\documentclass[12pt,,onecolumn,oneside,letterpaper]{article}
\usepackage{amsfonts, amssymb}
\usepackage{latexsym}
\usepackage{graphicx}
\usepackage{color}
\usepackage{setspace}
\usepackage[cmex10]{amsmath}
\usepackage{array}
\usepackage{mdwmath}
\usepackage{mdwtab}
\usepackage{cite}
\begin{document}
\title{A fast and light stream cipher for smartphones}
\author{G. Vidal, M. Baptista, H. Mancini}
%

%\institute{Enigmedia Corp., University of Aberdeen, University of Navarra}
\maketitle

\abstract{
We present a stream cipher based on a chaotic dynamical system. Using a chaotic trajectory sampled under certain rules in order to avoid any attempt to reconstruct the original one, we create a binary pseudo-random keystream that can only be exactly reproduced by someone that has fully knowledge of the communication system parameters formed by a transmitter and a receiver and sharing the same initial conditions. The plaintext is XOR’ed with the keystream creating the ciphertext, the encrypted message. This keystream passes the NIST's randomness test and has been implemented in a videoconference App for smartphones, in order to show the fast and light nature of the proposed encryption system.
} %end of abstract
\section{Introduction}
\label{intro}

Communications have gone into a revolution allowing faster trading, exchanging of ideas and the massive creation of social relations world wide. Unfortunately, security techniques have not been improved at the same speed. Current applications for mobile, tablets, smartcards, and other gadgets have considerabily increase the need for more bandwidth and CPU power, meaning that encryption and security techniques compete for these same resources. People have to choose between security or performance.

To create communication devices that are not only secure but have also good performance, the focus of research in security for these devices has shifted to the development of lighter and faster stream ciphers that require less CPU time. However, as less CPU consumption implies less algorithmic operations in the architecture of the stream cipher, these class of ciphers were considered to be insecure, weaker than the block ciphers \cite{Schneier}. For example, a bias on the well known stream cipher RC4 was discovered in 2001 \cite{RC4}, allowing to be attacked successfully in a polynomial time.

In this context the eStream project \cite{eStream} was born in 2004. Its aim was to give rise to a standardization of fast and lightweight stream ciphers. After receiving several proposals and years of analysis, only few of the original proposals were chosen to belong to the current official eStream portfolio. These cryptosystem are: HC-128, Rabbit, Salsa20/12 and SOSEMANUK in the software oriented category, and Grain v1, MICKEY 2.0 and Trivium in the hardware oriented category (see Ref. \cite{eStream} and the reference therein).

In this manuscript, we propose a new approach for building stream ciphers based on Chaos Theory and its well known confusion and diffusion properties \cite{Kocarev}. We propose a new fast and light stream cipher, named Enigmedia, which is based on a hyperchaotic dynamical system, a codifying method with a whitening technique and a nonlinear transformation in order to obtain a pseudo-random keystream sequence with high performance. The message is XORed in one side with the keystream, obtaining the ciphertext. As the method is deterministic, this keystream can be generated in the other communication side. Then, by XORing the same keystream one can recover the original message.

One of the main feature of this cryptosystem is that the initial conditions used by both transmitter and receiver do not need to be equal, allowing the system to posses a random secret key. However, in order to minimize the set-up call time, we impose that the initial conditions for receiver and transmitter are equal. It implies that we can not use another feature of this system: if an eavesdroper tries to intercept the communication, both receiver and transmitter become aware of that, as it happens in quantum cryptography \cite{VBM}.

In this work, we have shown that for the creation of a good pseudo-random sequence, only a small number of integrations of the chaotic systems are needed. That is the reason of why the Enigmedia cipher is fast and light. This result is a consequence of the fact that chaotic systems have an exponentially fast decay of the correlation. The used system has a particular large exponential decay, since it is a higher dimensional chaotic system with two positive Lyapunov exponents. Additionally, two Lyapunov exponents make impossible to detect signatures such as universabilities or initial condition accotation.

We use the mutual information to show decay of correlation of two points of the chaotic trajectory separated by a small time interval. This time interval is used to sample the chaotic trajectory from which we calculate the keystream. We show by using mutual information that also the keystream behaves as on independent binary sequence. If the mutual information between points (or samples) in a time-series (or binary sequence) is zero, that means that that these time-series (or binary sequence) behave as an independent random process.

To improve further the speed of our cipher, we expand the keystream created by the high dimensional chaotic system, using a nonlinear map. These longer sequences were shown also to pass the NIST's tests, which assure their random and independence nature.

Furthermore, we present a benchmark of the performance between our encryption system and actual \emph{de facto} standards in block and stream ciphers, i.e., AES-128 and RC-4. Finally, we show how our implementation for smartphones passes NIST's randomness tests showing the high quality of the keystream.

\section{Communication Scheme and Arquitecture}
\label{Sec:Arquitecture}

\begin{figure}
 \begin{center}
  \includegraphics[width=\textwidth]{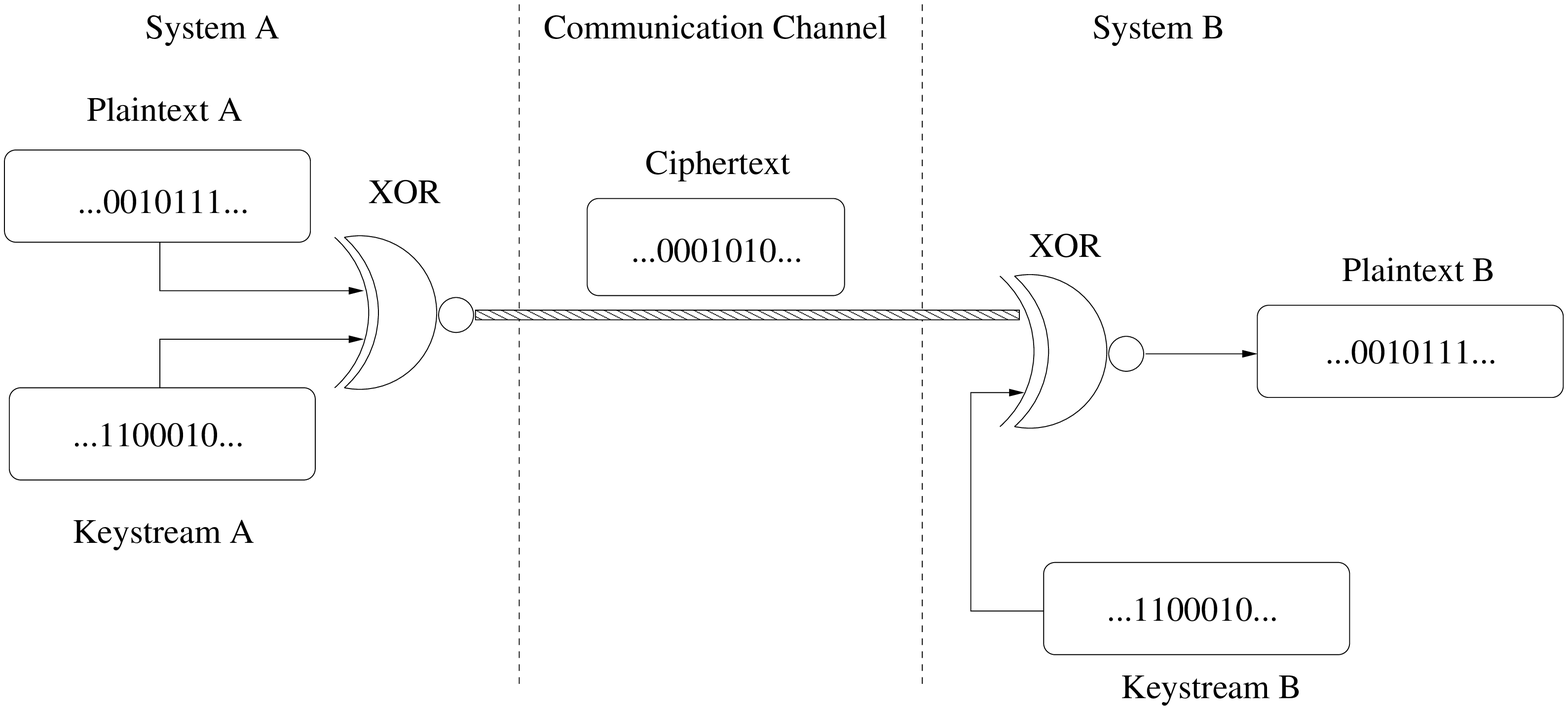}
 \end{center}
  \caption{Communication Scheme. Both communication extremes must share the same key in order to generate the same keystream for recovering the original plaintext.}
  \label{Comscheme}
\end{figure}

As a stream cipher, Enigmedia is a symmetric cryptographical system. It means that emitter (Alice, A) and receiver (Bob, B) must share the same key in order to encrypt/decrypt the message sent through the communication channel. With these key or seed, both A and B generates the same keystream. XORing this keystream with the plaintext/ciphertext one obtain the ciphertext/plaintext as it is shown in Figure \ref{Comscheme}.

Note that it is important not to repeat the same key because if an eavesdrop (Eve, E) identifies the keystream, it could use this information to recover a message. In order to avoid this, we use a true random number generator \cite{Stallings} to generate our initial conditions for the chaotic trajectory. These systems use a nondeterministic source of randomness and by using them we assure a key randomly chosen.

As all symmetric systems, it is required a secure channel to exchange the key between Alice and Bob \cite{Handbook}. In our implementation for secure videoconference, we use a Public Key Infrastructure (PKI) and electronic certificates for assuring authentication and secure key exchange via assymmetric cryptography techniques.

It is important to remark that this cryptosystem is implemented in the higher layers on the OSI (Open System Interconnection) model \cite{OSI}, this implies that:

I) The encryption does not depend on the physical medium of the communication channel, i.e., radiowaves, electrical signals, optical fiber, etc.

II) Data transmission is robust to noise because we use other protocols to assure quality of service and check errors. When these errors are produced, information packets can be retransmitted or simply discarded.

\section{The Dynamical System}

The dynamical system arises from a pattern formation model in a convective pattern experiment \cite{PRL_94}. The equation set models how the dynamics of pattern evolves according to a control parameter,i.e., the heating from below in a convective cell. The equation set is the following:

\begin{eqnarray}\nonumber \label{eqs_d4}
\dot{x} &=& y\\ \nonumber
\dot{y} &=& \mu x+x(a(x^2+z^2)+bz^2)\\
\dot{z} &=& w\\ \nonumber
\dot{w} &=& \mu z+z(a(x^2+z^2)+bx^2) \nonumber
\end{eqnarray}

We use $a<0, b>0, \mu>0$ assuring the chaotic behaviour of the trajectories \cite{IJBC_D4}.

The dynamical system used has some properties which makes it specially interesting in cryptographical applications \cite{VBM}. Most significant are hiperchaoticity in the sense of Rössler \cite{Rossler_HC} and the inner symmetries that appear in the equation system. Another important property is that it is not an attractor in the sense of Milnor \cite{Milnor}, it is a non-attractive set with a riddled basin \cite{Riddled_IJBC}. In further sections we describe why these properties are important for security.

\section{Generating Sequences}

Here we explain how a chaotic system such as in Eq.(\ref{eqs_d4}) can be used for obtaining a pseudo-random sequence.

\begin{figure}
 \begin{center}
  \includegraphics[width=\textwidth]{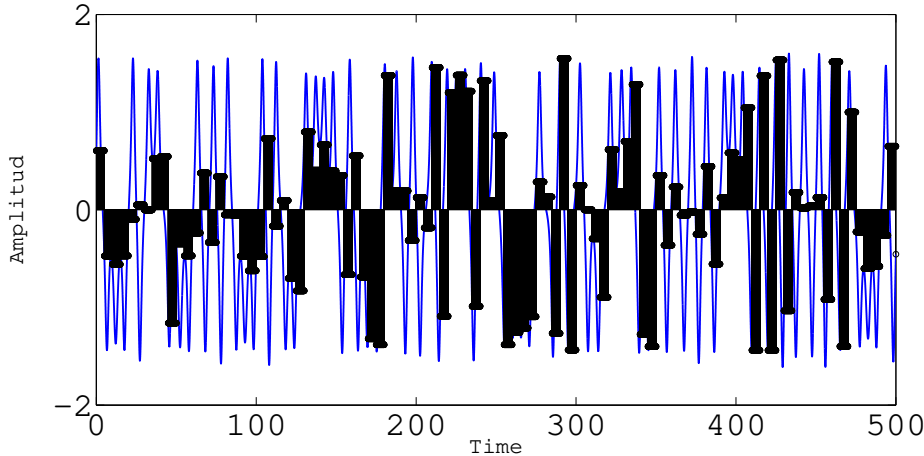}
 \end{center}
  \caption{We represent the sampling process. The blue line plots a trajectory and dark steps show the samples ``holded'' for some time for illustrating purpouses.}
  \label{Sampling_Trajectory}
\end{figure}

First of all, assume that an initial condition is randomly chosen as explained in Sec. \ref{Sec:Arquitecture}. From  this seed we compute the trajectory using a numerical method (for example Runge-Kutta 4th order). Then, this trajectory is generated considering some integration time steps $\delta t$. We sample this trajectory considering a certain number of steps $m$. Let us consider for a time interval $T_0 = m\delta t$. If we choose properly this number of steps, temporal correlation between any two points of the sampled trajectory will be low. Fig. \ref{Sampling_Trajectory} represents this sampling process.

A special characteristic of chaotic systems is that the time $T_0$ for the correlation to decay to neglegible values is finite and short. If the system being used is hyperchaotic, such as in Eq.(\ref{eqs_d4}) the speed for the correlation decay is advantadgeously fast. In order to assure that, we calculate the Mutual Information between 2 points separated by a time interval $T_0$ according to Abarbanel \cite{Abarbanel}.

We use Mutual Information because it is a measure of the amount of information that one random variable contains about another random variable \cite{Cover}. Consider the random variables $X$ and $Y$ with possible measurements $x\in\mathcal{X}$ and $y \in \mathcal{Y}$ and a joint probability mass function (p.m.f.) $p(x,y)$ with marginal p.m.f. $p(x)$ and $p(y)$, then mutual information $I(X,Y)$ is defined by:

\begin{equation}\label{eq:MI}
I(X;Y) = \sum_{x \in \mathcal{X}} \sum_{y \in \mathcal{Y}} p(x,y)\log \frac{p(x,y)}{p(x)p(y)}
\end{equation}
where the $\log$ is in base 2. Note that a mutual information of zero means that no information about $X$ can be learnt by the sole observation of $Y$.

In order to place this abstract definition to our work, given a trajectory represented by $s(t)$, we represent it by a discrete time-series $s(n)$, where $n$ represents the number of integration steps. The set of observations as $X$ is replaced by $s(n)$, whereas the set $Y$ is replaced by the ``delayed trajectories'' $s(n+T)$. The average mutual information between observations at $n$ and $n+T$,i.e., the average amount of information about $s(n+T)$ when we observe $s(n)$ is then:
\begin{equation}\label{eq:average_MI}
I(T) = \sum^{N}_{n=1} p(s(n),s(n+T))\log \frac{p(s(n),s(n+T))}{p(s(n))p(s(n+T))}
\end{equation}
with $I(T)\geq0$, according to \cite{Abarbanel}.

\begin{figure}
 \begin{center}
  \includegraphics[width=\textwidth]{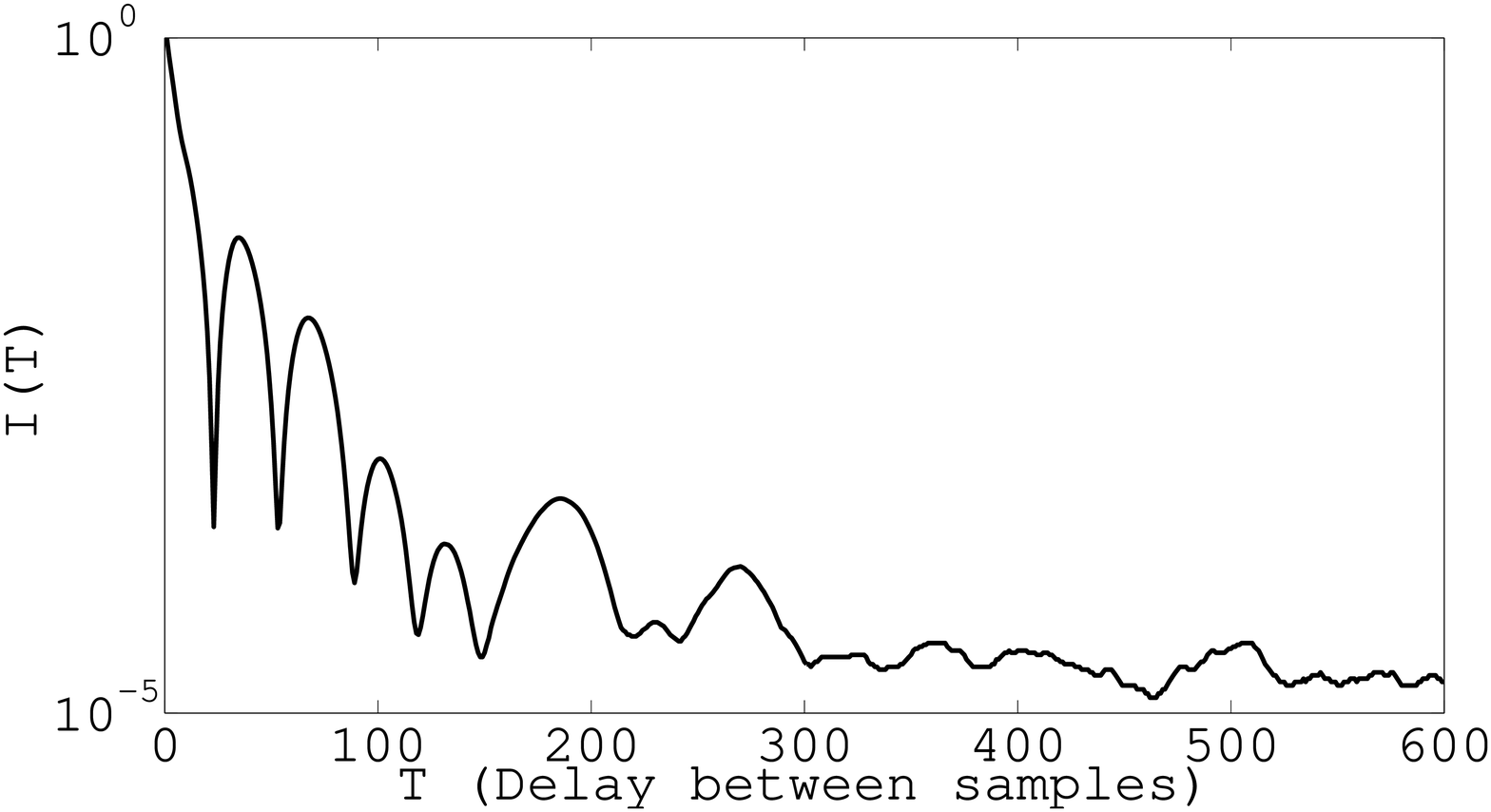}
 \end{center}
  \caption{Mutual Information for the delayed system. This result shows how the system loose its autocorrelation with time. This plot provide us with information about how to sample the system.}
  \label{MI_no_sampled}
\end{figure}

For computing these probabilities we create an histogram for the different possible values of the trajectory. We generate one of these histograms per each $n$ and $n+T$ in order to compute the joint distribution. These histograms accumulate the values of the trajectories in $n$ and $n+T$. In this case, we simulate for several different trajectories generated from different initial conditions randomly chosen. The value of $I(T)$ is shown in Fig. \ref{MI_no_sampled} as a function of $T$. Notice that for small values of $T$, $I(T)$ can already be very low.

By sampling the trajectory for a $T_0$ with $I(T_0) \approx 0$ the series of sampled values are reasonably uncorrelated to each other, but the distribution of the values will not be uniform. This happens because as in most of the chaotic systems, there are certain regions in the phase space more ``visited'' than others.

\begin{figure}
 \begin{center}
  \includegraphics[width=\textwidth]{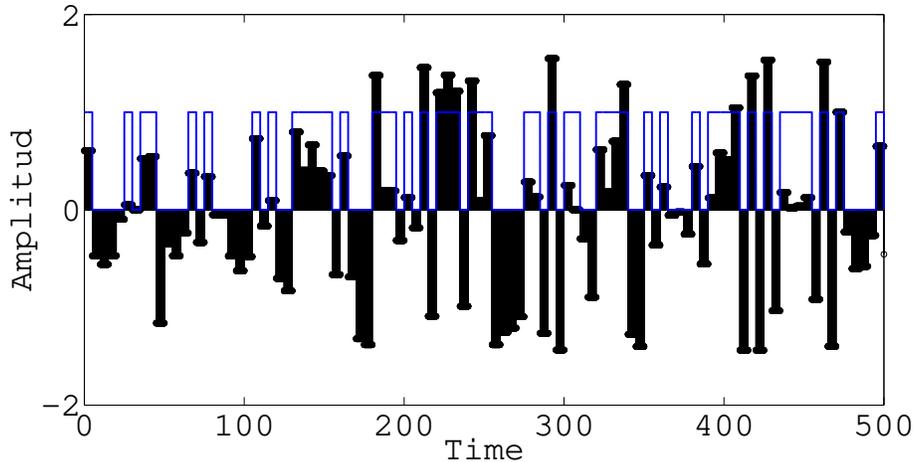}
 \end{center}
  \caption{We represent the whitening process. The blue line representing the symbolic values of the keystream shows a possible transformation in order to obtain an equally distributed amount of samples around the symmetry axis from the sampled values (in dark color).}
  \label{Binaryzing_Trajectory}
\end{figure}

In order to obtain a uniform distribution and decrease further the correlation, we apply a whitening process for destroying the spatial correlation. It consists on binarizing the variable sampled respect to the symmetry axis. Choosing the variable $x$ as to be the one generating the keystream, we consider the trajectory can be encoded by a $'0'$ if $x<0$, and encoded by $'1'$ if $x>0$. Due to the inner symmetries of the system in Eq.(\ref{eqs_d4}), the distribution of the trajectory points will be the same in each side (when the number of samples are big enough according to the central limit theorem). The fundamental idea behind the whitening process is that we are creating an approximation to a Markov's partition of the sampled trajectory into a symbolic sequence, as it is shown in Fig. \ref{Binaryzing_Trajectory}.

Assuming that we are computing the trajectories in double format,i.e., 32 bit precision, we can model this process as a boolean function which relates a 4 variable system of 32 bits each to one information bit, i.e., $F:\ \mathbb{F}_2^{4\cdot 32} \rightarrow \mathbb{F}_2^2$ applied over the samples. 

Therefore, we obtain the binary pseudorandom sequence, i.e., the keystream.

\begin{figure}
 \begin{center}
  \includegraphics[width=\textwidth]{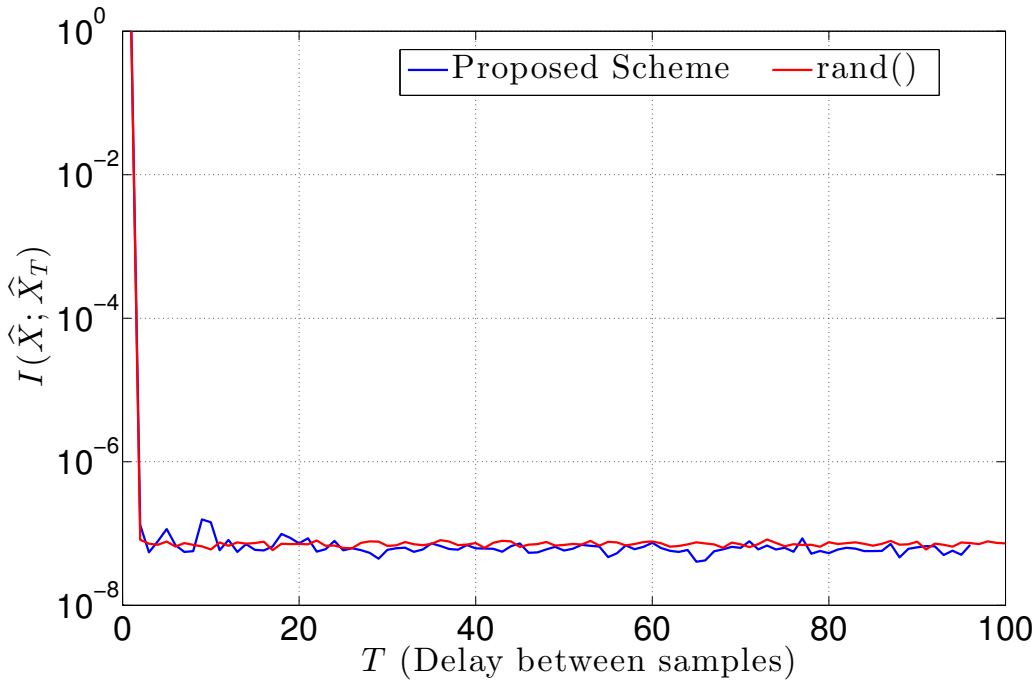}
 \end{center}
  \caption{Mutual Information for sampled system. Also we compare the keystream generated by our proposed scheme with rand() C function.}
  \label{MI_sampled}
\end{figure}

If we calculate again the mutual information for this keystream, interpreting $s(n)$ in Eq.(\ref{eq:average_MI}) to represent the binary symbol $\hat{X}$ and $s(n+T)$ the binary symbol shifted forward in time $\hat{X_T}$, and we compare it with the mutual information of binary sequence generated by rand() function in C with equiprobable 0s and 1s, we show in Fig. \ref{MI_sampled} that the mutual information of the binary keystream to the one obtained from the random binary sequence as a function of $T$. The similarities between both curves asserts the good statistical properties. We have chosen this function due to its very good statistical properties of the proposed system. Notice, however, that this function has bad cryptographical properties because numbers generated by rand() have a comparatively short cycle, and the numbers can be predictable.

\section{Experiments and Implementation}

In cryptography, it is important to have both speed and security, otherwise there is no reason to use a new system. The current standard AES \cite{AES} has been tested since 90's and its special designed hardware chipsets are common in high-standing handheld devices to increase its efficiency.

For improving the speed of the proposed cipher, we have used an expanding nonlinear function based on the Baker's Transformation \cite{Schuster} to expand the binary sequence generated with the system of Eq.(\ref{eqs_d4}). As a simple example of the concept, let be two sequences such as $(00,11)$. We use a transformation or rule to expand them, for example, $(101010-001101)$. Then we use another transformation to wrap them up $(1011-1100)$. The final sequence is larger than the initial ones, created by mixing and stretching the streams. Notice that by using a folding operation we obtain a final sequence that is not a one-to-one mapping from the initial ones, for more details see \cite{Patent}.

There are other stream ciphers which use similar techniques, such as Rabbit stream cipher \cite{Rabbit}. A set of libraries for implementing the operations and numerical methods in order to assure portability between devices while keeping the speed have been developed.

In order to assure the portability of the new cryptosystem, it has to run the same algorithm in different devices. We have used the following devices and processors:
\begin{enumerate}
 \item Samsung Galaxy S - ARM Cortex-A8 @ 1 GHz
 \item Android TouchBook - ARMv7 OMAP3 @ 720 MHz
 \item Notebook EeePC Asus - AMD C-60 APU @ 800 MHz
 \item ThinkPad Lenovo - Intel Core 2 i3 @ 2 GHz
\end{enumerate}

Benchmark shows how many CPU clock cycles per byte encrypted $(\frac{cycles}{byte})$ takes to implement Enigmedia in different architectures. In this implementation, we do not parallelize processes and operations, and we do not use SIMD operations, except in encryption rate speed tests done for Samsung Galaxy S where we use Neon set operations. Results obtained are shown in Table \ref{Benchmark} where Enigmedia is compared against AES-128 and RC4 with \emph{OpenSSL} implementation, which enables the use of dedicated hardware in most of laptops to improve the performance of AES and RC4.

It is important to remark that even in this situation Enigmedia uses at least one order of magnitude less CPU clock cycles than those ciphers

\begin{tabular}{|c| c| c|c|}
\hline & & &\\
 Processor &  Enigmedia Rate & AES Rate & RC4 Rate \\
  \hline & & & \\
 ARM A8 1 GHz & 2.33 $\frac{cycles}{byte}$ & *** & ***\\
  \hline & & & \\
ARM A8 1 GHz & 11.00 $\frac{cycles}{byte}$ & *** & ***\\
  \hline & & & \\
 ARMv7 720 MHz & 4.80 $\frac{cycles}{byte}$ & 55.28 $\frac{cycles}{byte}$ & 20.79 $\frac{cycles}{byte}$ \\
 \hline & & & \\
 AMD C-60 800 MHz & 2.95 $\frac{cycles}{byte}$ & 49.95 $\frac{cycles}{byte}$ & 7.56 $\frac{cycles}{byte}$\\
 \hline & & &\\
 Intel i3 2GHz & 1.77 $\frac{cycles}{byte}$ & 32.26 $\frac{cycles}{byte}$ & 10.92 $\frac{cycles}{byte}$\\
 \hline & & &
 \label{Benchmark}
\end{tabular}
\smallskip

Futhermore, we have checked the randomness of our expanded and wraped binary keystream with NIST's tests \cite{NIST}, which is a battery of statistical tests to detect non-randomness in binary sequences constructed using pseudo-random number generators. Some of these tests check the characteristic functions, patterns and rare events which should appear in random sequences. We use them as they are a necessary, but not sufficient condition, to assure randomness. These tests focus on a variety of different types of non-randomness that could exist in a sequence. The 15 tests are:

\begin{enumerate}
 \item The Frequency (Monobit) Test
 \item Frequency Test within a Block
 \item The Runs Test
 \item Tests for the Longest-Run-of-Ones in a Block
 \item The Binary Matrix Rank Test
 \item The Discrete Fourier Transform (Spectral) Test
 \item The Non-overlapping Template Matching Test
 \item The Overlapping Template Matching Test
 \item Maurer's "Universal Statistical" Test
 \item The Linear Complexity Test
 \item The Serial Test
 \item The Approximate Entropy Test
 \item The Cumulative Sums (Cusums) Test
 \item The Random Excursions Test
 \item The Random Excursions Variant Test
\end{enumerate}

A number of tests in the test suite have the standard gaussian normal and the $\chi^2$, or Pearson's distribution, as reference distributions. $\chi^2$ is used to compare the goodness-of-fit of the observed frequencies of a sample measure to the corresponding expected frequencies of the hypothesized distribution.

If the sequence under test is in fact non-random, the calculated test statistic will fall in extreme regions of the reference distribution. A $P$-value summarizes the value of these ``tails'' and helps us to know if a long sequence is random or not. We can calculate this $P$-value as given by:
\begin{equation}\label{eq:P_value}
P\mathrm{-value} = igamc\left(\frac{9}{2},\frac{\chi^2}{2}\right) = \frac{\Gamma(\frac{9}{2},\frac{\chi^2}{2})}{\Gamma (\frac{9}{2})}=\frac{1}{\Gamma(\frac{9}{2})}\int^{\infty}_{\frac{\chi^2}{2}} e^{-t}t^{\frac{9}{2}-1}dt
\end{equation}
where $igamc$ is know as the \emph{incomplete gamma function}. As a rule of thumb, if $P$-value $\geq 0.0001$ the distribution is considered uniform.

\begin{figure}
 \begin{center}
  \includegraphics[width=\textwidth]{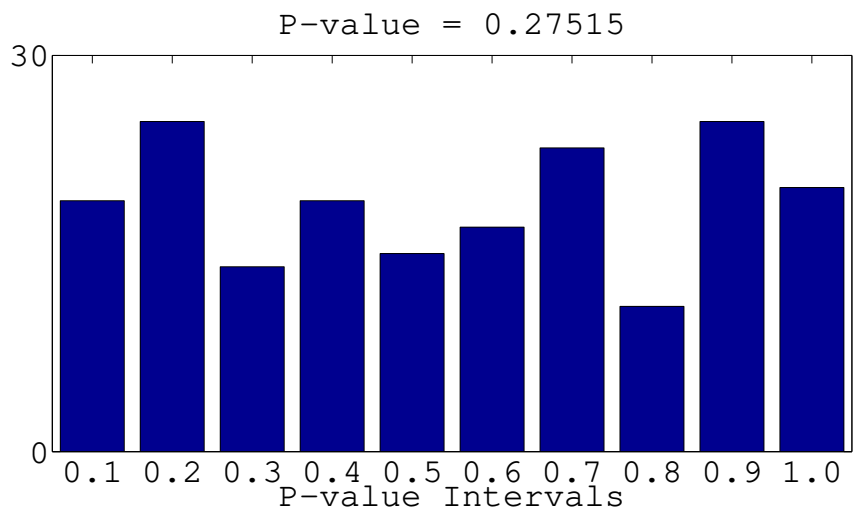}
 \end{center}
  \caption{NIST's tests results for 800 sequences of 1Mb each. The $P$-value is obtained considering the results of all these tests. As the $P$-value $\geq 0.0001$ we can consider that the sequences are pseudorandom.}
  \label{NIST_tests}
\end{figure}

The distribution of $P$-values is examined to ensure uniformity. This may be visually illustrated using a histogram, whereby, the interval between 0 and 1 is divided into 10 sub-intervals, and the $P$-values that lie within each sub-interval are counted and displayed. We show in Fig. \ref{NIST_tests} the $P$-value for the all the tests. As the keystream are supposed to be random, then strange events should appear. Uniformity on these results is shown by determination of a $P$-value corresponding to the Goodness-of-Fit Distributional Test over all tests. This is accomplished by computing:
\begin{equation}\label{eq:chi_square}
\chi^2=\sum_{i=0}^1\frac{\left(F_i - \frac{s}{10}\right)^2}{\frac{s}{10}}
\end{equation}
where $F_i$ is the number of $P$-values in sub-interval $i$, and $s$ is the sample size. Then the average $P$-value is calculated appling Eq.(\ref{eq:P_value}),  providing the value 0.27515. This value comes from statistical tests over 800Mb from 1 seed of 252 bits, which is the keylenght of the implementation. We have applied same tests in different sequences obtained from random initial conditions obtaining similar results, and validating the system.

In order to test efficiency of the proposed cipher Enigmedia implemented also a videoconference application, which helps to measure CPU consumption, speed, and how sensible are the encryption method to the technical characteristics of the devices.

\section{Conclusions}

We present a light and fast cipher based on chaos. The idea is to create a binary keystream by encoding a sampled chaotic trajectory every time interval $T_0$. This time is chosen in order to produce a sampled trajectory that has very low correlation and a keystream that has similar behaviour as a random independent process. The time interval is the shortest time for which mutual information of the sampled trajectory decays to close to zero values.

For a real application of our own theoretical ideas, we improve the speed of the cipher by expanding the keystream using the help of a nonlinear transformation. This final keystream has shown to pass the NIST's tests to assure its random and independent nature.

Finally, it is known that having a very good statistical properties is a necessary condition for a system to be secure, but it is far from been a sufficient condition \cite{Handbook}. For example the linear congruential generator (the \emph{rand()} function provided by C) is not secure due to the linearity of the generation process \cite{Handbook}. However, since our function posses a high nonlinear component, we believe that our function is also robust against the state of the art attacks on stream ciphers (as the algebraic and correlation attacks \cite{Wu, Hell}). Moreover, the fact that the states are updated by means of several realizations of an ODE solver, makes us believe that a description of the system in terms of near-linear boolean functions is not possible.

\section{Acknowledgements}

H.M. wants to thanks financial support from Spanish Ministry of Science and Technology under Contract NO. FIS2011-24642.


\begin{thebibliography}{99}

\bibitem{AES} Joan Daemen and Vincent Rijmen, "The Design of Rijndael: AES The Advanced Encryption Standard." Springer, 2002.
\bibitem{RC4} G. Paul and S. Maitra, "RC4 Stream Cipher and Its Variants", B\&N, 2011.
\bibitem{eStream} eStream Project, info available at \emph{http://www.ecrypt.eu.org/stream/}.
\bibitem{Kocarev} L. Kocarev, "Chaos-based cryptography: a brief overview," Circuits and Systems Magazine, IEEE , vol.1, no.3, 2001
\bibitem{Lyapunov} O.E. Rossler, "An equation for Hyperchaos",\emph{Phys. Lett.}, 71A, 1979.
\bibitem{VBM} G. Vidal, M. S. Baptista and H. Mancini,  "Fundamentals of a Classical Chaos-Based Cryptosystem with some Quantum Cryptography Features". \emph{I. J. Bifurcation and Chaos}, 22 (10), 2012.
\bibitem{NIST} NIST randomness test suit, available at \emph{http://csrc.nist.gov/groups/ST/}.
\bibitem{Handbook} A. J. Menezes, P. C. van Oorschot and S. A. Vanstone, "Handbook of Applied Cryptography", CRC Press, 1997.
\bibitem{Wu} Hongjun Wu, "Cryptanalysis and Design of Stream Ciphers", PhD dissertation, Katholieke Universiteit Leuven, 2008.
\bibitem{Hell} Martin Hell, "On the Design and Analysis of Stream Ciphers", PhD dissertation, Lund University, 2007.
\bibitem{OSI}  H. Zimmermann ``OSI Reference Model--The ISO Model of Architecture for Open Systems Interconnection,'' {\it IEEE Transactions on Communications} {\bf 28}, pp. 425--432, 1980.
\bibitem{Rossler_HC} O.E. R\"ossler [1979] ``An equation for hyperchaos,'' {\it Physics Letters} {\bf 71A}, pp. 155--157.
\bibitem{PRL_94} Ondar\c{c}uhu, T. Mindlin, G., Mancini, H. L. \& P\'erez-Garc\'{\i}a C. [1984] ``The chaotic evolution of patterns in B\'enard-Marangoni convection with
  square symmetry,'' {\it J. Phys.: Condens. Matter A} {\bf 6} pp. 427--432.
\bibitem{Milnor} Milnor, J. [1985] ``On the concept of attractor,'', \textit{Commum. Math. Phys.} \textbf{99}, pp. 177--195.
\bibitem{Riddled_IJBC} Alexander, J. C., Kan, I., Yorke, J. A., {\it et al.} [1992] ``Riddled basins,'' {\it
    Int. J.  Bif. Chaos} {\bf 2}, pp. 795--813.
\bibitem{Schneier} Kelsey, J., Schneier, B., Hall, C., ``Cryptanalytic Attacks on Pseudorandom Number Generators,'' {\it Proc. Fast Software Encryption}, 1998.
\bibitem{Abarbanel} H. Abarbanel R. Brown, J.J. Sidorowich, L.S. Tsimring, ``The analysis of observed chaotic data in physical systems'' {\it Rev.Mod.Physics},Vol. 65,No. 4, 1993.
\bibitem{Stallings} W. Stallings, {\it Cryptography and Network Security}, Pearson, 2011.
\bibitem{Cover} T. Cover, J. Thomas, {\it Elements of Information Theory}, Wiley, 1991.
\bibitem{Schuster} Schuster, H. {\it Deterministic Chaos: An Introduction} Physik-Verlag, 1984.
\bibitem{Rabbit} M. Boesgaard, M. Vesterager, T. Pedersen, J. Christiansen, O. Scavenius. ``Rabbit: A High-Performance Stream Cipher'', {\it Proc. FSE 2003}, Springer LNCS 2887, 2003.
\bibitem{NIST_doc} A. Rukhin {\it et al.}, ``A Statistical Test Suite for Random and Pseudorandom Number Generators for Cryptographic Applications'', \emph{NIST Special Publication 800-22A}, 2010.
\bibitem{IJBC_D4} G. Vidal, \& H. Mancini [2009] ``Hyperchaotic Synchronization under square symmetry,'' {\it  Inter. Jour. of Bif. and Chaos} {\bf 19}, pp. 719--726.
\bibitem{Patent} European Patent - EP12382201, Patent Pending
  
\end{thebibliography}
\end{document}